\documentclass[11pt]{article}
\usepackage{hyperref}
\pdfoutput=1
\begin{document}
\title{Ripples and Shear Bands in \\ Plowed Granular Media}
\author{Nick Gravish \\
\\\vspace{6pt} School of Physics, Georgia Tech}

\maketitle
\begin{abstract}
fluid dynamics videos.
\end{abstract}

Video
\href{http://ecommons.library.cornell.edu/bitstream/1813/14093/3/ripples_dfd09_mpeg2.mpg}{(mgpeg-2)}
\href{http://ecommons.library.cornell.edu/bitstream/1813/14093/2/ripples_dfd09_mpeg1.mpg}{(mgpeg-1)}
\section{Summary}

Monodisperse packings of dry, air-fluidized granular media typically exist between volume fractions from $\Phi$= 0.585 to 0.64. We demonstrate that the dynamics of granular drag are sensitive to volume fraction $\Phi$ and their exists a  transition in the drag force and material deformation from smooth to oscillatory at a critical volume fraction $\Phi_{c}=0.605$. By dragging a submerged steel plate (3.81 cm width, 6.98 cm depth) through $300 \mu m$ glass beads prepared at volume fractions between 0.585 to 0.635 we find that below $\Phi_{c}$ the media deformation is smooth and non-localized while above $\Phi_{c}$ media fails along distinct shear bands. At high $\Phi$ the generation of these shear bands is periodic resulting in the ripples on the surface. Work funded by The Burroughs Wellcome Fund and the Army Research Lab MAST CTA

\end{document}